# Secure Virtual Mobile Small Cells: A Stepping Stone Towards 6G

J. Rodriguez[1,11], G. P. Koudouridis[2], X. Gelabert[2], M. Tayyab[3,9], R. Bassoli[4], F. H.P. Fitzek[4], R. Torre[4], R. Abd-Alhameed[5], M. Sajedin[1,10], I. Elfergani[1], S. Irum[6,4], G. Schulte[6], P. Diogo[7], F. Marzouk[7,10], M. de Ree[1,11], G. Mantas[1], I. Politis[8]

[1]Instituto de Telecomunicações, Portugal, [2]HUAWEI Technologies Sweden R&D, Sweden, [3]HUAWEI Technologies Oy Co Ltd, Finland, [4]Technische Universität Dresden, Germany, [5]University of Bradford, U.K., [6]Acticom Gmbh, Germany, [7]PROEF SGPS SA, Portugal, [8]University of Patras, Greece, [9]Aalto University, Finland, [10]Universidade de Aveiro, Portugal, [11]University of South Wales, UK

*Abstract*—As 5th Generation research reaches the twilight, the research community must go beyond 5G and look towards the 2030 connectivity landscape, namely 6G. In this context, this work takes a step towards the 6G vision by proposing a next generation communication platform, which aims to extend the rigid coverage area of fixed deployment networks by considering virtual mobile small cells (MSC) that are created on demand. Relying on emerging computing paradigms such as NFV (Network Function Virtualization) and SDN (Software Defined Networking), these cells can harness radio and networking capability locally reducing protocol signaling latency and overhead. These MSCs constitute an intelligent pool of networking resources that can collaborate to form a wireless network of MSCs providing a communication platform for localized, ubiquitous and reliable connectivity. The technology enablers for implementing the MSC concept are also addressed in terms of virtualization, lightweight wireless security, and energy efficient RF. The benefits of the MSC architecture towards reliable and efficient cell offloading are demonstrated as a use-case.

## I. Introduction

It is widely accepted that the 6G drive aims at promoting digital inclusion and accessibility, as well as unlocking economic value and opportunities in rural communities [1]. Harnessing on the services already offered by 5G technology, 6G aims to integrate an even richer set of services, that includes virtual and augmented reality, and autonomous vehicles, among others. This will require disruptive architectures that can build on 5G to deliver market relevant solutions. In this context, the authors are involved in the MCSA-ETN SECRET initiative [2], a European training network that aims to take a step towards the 6G vision by proposing a next generation communication platform that extends the coverage area of 5G users. The SECRET platform enables mobile devices to be perceived by the mobile network as a pool of available virtual resources that have computing, storage and networking capability; i.e. end-users can become autonomous small cells inheriting radio resource management capabilities, referred as Mobile Small Cells (MSCs). The SECRET connectivity landscape envisages a wireless network of MSCs covering the urban landscape, which are virtual in nature and set-up on demand, providing a basis for localized small cell connectivity. The platform is controlled by a distributed SDN (software defined network) architecture enabling opportunities such as network slicing, cloud computing, as well as lowering the total cost of ownership (TCO) [3] for Mobile Network Operators (MNOs) that are integral to the 6G philosophy.

In this context, SECRET explores proven technology paradigms based, among others, on Network Coded Cooperation (NCC) and Virtualization. Whilst virtualization is implemented based on mature computing tools such as Software Defined Networking (SDN) and Network Function Virtualization (NFV), NCC has been proven to be essential for current mobile paradigms [4]. NCC allows entities to interact and collaborate to improve transmitted signal quality, while making effective use of the system bandwidth. These two paradigms were merged to provide an integrated architecture that builds on 5G with potential B5G and 6G requirements on communication, computation, caching and control optimization that can enable TCO efficiency gains and meet latency and reliability targets. This integrated SECRET approach raises significant challenges in terms of enabling virtualization, mobility, wireless security, and energy efficient RF that are coherent with the 6G technology challenges.

The key technologies and standardization trends to addressing these challenges and implementing virtual MSCs, as demonstrated through efficient and reliable cell offloading, are the focus of this paper.

## II. SECRET Vision: Mobile Small Cells for B5G

The SECRET reference scenario envisages dynamically created MSCs covering a limited geographical area, which are virtual in nature. They can be setup on demand, anywhere, anytime, and on any 6G-enabled mobile (or fixed) device based on network conditions and device capabilities. From the end-user perspective, these MSCs support a plethora of

This project has received funding from the European Union´s H2020 research and innovation program under grant agreement H2020-MCSA-ITN- 2016-SECRET 722424 [2].

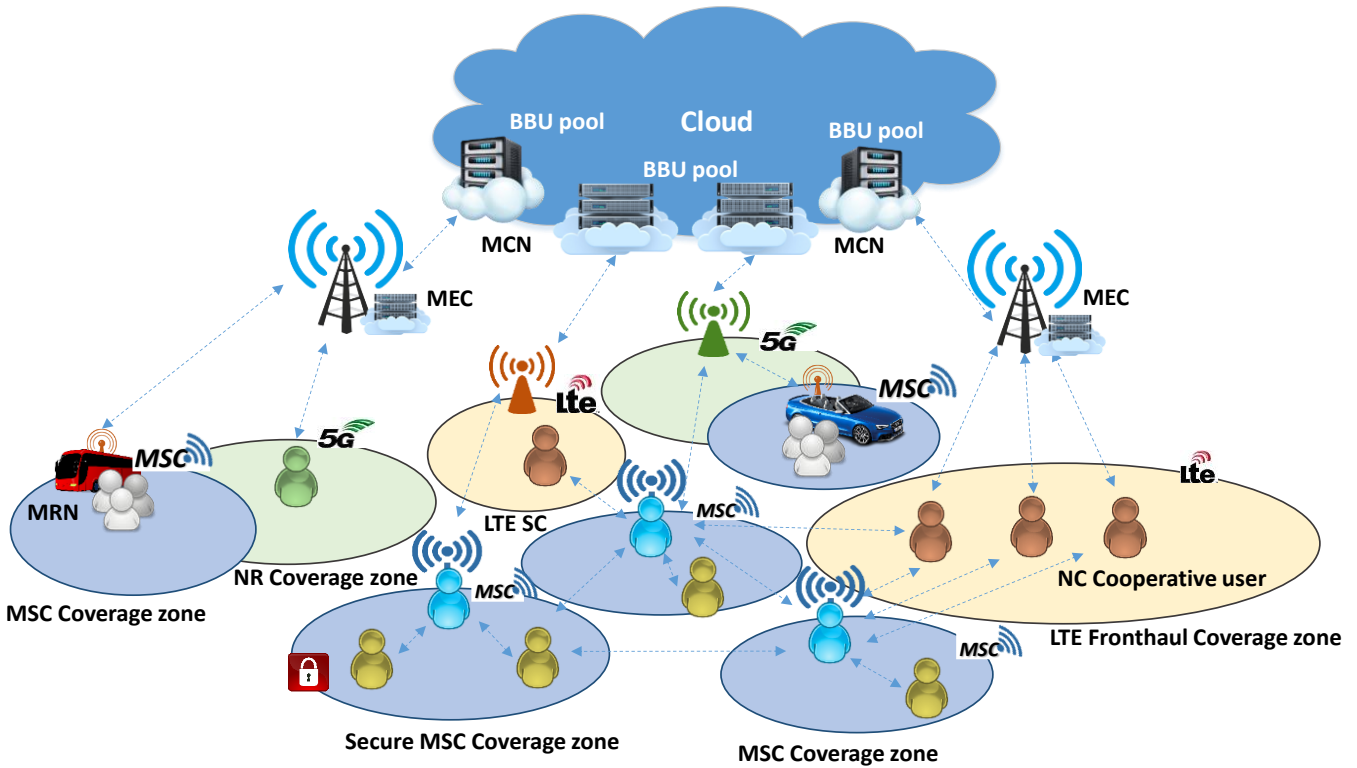

Figure 1. Overall SECRET reference scenario. BBU: Baseband units, LTE: Long-term Evolution, MCH: Mobile Cell-Head, MCN: Mobile Core Network, MEC: Mobile Edge Computing, MSC: Mobile Small Cell, NC: Network-coding, NR: New Radio, SC: Small Cell, UE: User Equipment.

potential 6G services at low cost, and with reduced impact on mobile battery lifetime. Each MSC is controlled by a mobile cell head (MCH), which is a mobile device within the identified cluster set that is nominated to become the local radio manager. The MCH controls and maintains the set of active users in the coverage area of the cell. As illustrated in Fig. 1, the MCHs cooperate with MCHs of other MSCs to form a "wireless-network-of-small-cells" that can have several gateways to either the mobile core network (MCN) through the infrastructure-based network (4G macro/5G NR), or other User Equipment (UEs) e.g., for cell offloading.

A main characteristic of the MSCs is that they utilize wireless fronthaul and backhaul connections from the Cloud Radio Access Network (CRAN), via the mobile edge computing (MEC) node to the MSC, avoiding the need for mobility-bounded and costly fiber-optic deployment. The wireless fronthaul link can support multiple radio access technologies (MRAT) and bands, including LTE-A and 5G NR. Another characteristic of MSCs is that they appear to the mobile devices as fully-fledged base stations (BSs), such as enhanced NodeBs (eNBs), providing local access via backhaul connectivity to a pool of baseband processing units (BBUs). The BBU pool is a virtualized cluster, composed of multiple processors and computing nodes performing baseband processing functionalities that, in practice, can be distributed throughout the network. Hence, MCN, MEC servers and the MSCs can all be considered as computing nodes. The communication from the MCN down to the MSCs is based on optimal virtualization of the fronthaul, backhaul and BBU resources adapting to the UE requirements, service level agreements and computational load.

To this end, the SECRET scenario targets a virtualization of radio access networks (RANs) for MSCs that are controlled by a distributed SDN architecture. Such an architecture enables adaptable deployment of computation, communication, caching and control (4C) solutions to include mobile small cells, allowing 4C to reach new limits in terms of coverage, throughput, latency and reliability. Incorporating 4C enabling technology, SECRET has the potential to deploy "widespread" functional split and network slicing to support 6G use-cases such Ultra-Reliable-Low Latency-Communications (URLLC) or critical Machine-Type-Communications (cMTC); stretches the computing coverage for task off-loading services from the edge cloud to devices in the near vicinity, lending to the term coined MIST computing [5]. This will provide a new dimension towards introducing further latency saving outside communication cost; as well as addressing cross-generation requirements in terms of reducing cost of ownership for mobile operators. Virtualization will allow the mobile device to become a pool of networking and computing resources that can be shared between operators and services.

The implementation of the MSC concept and networking has raised significant challenges in terms of device virtualization, wireless security and energy efficient RF, that can be depicted by the SECRET reference scenario. To achieve a flexible network where energy, spectral, and network usage efficiency meet the targets of B5G/6G networks, requires at least: (a) SDN-based and NFV-based CRAN supporting virtualization and softwarization; (b) energy-efficient (EE) handover (HO) control mechanisms that cope with the high mobility of connected devices acting as cell-heads; (c) decentralized wireless security framework that establishes cryptographic keying material to secure cooperation connections among devices; and (d) "green" PAs aiming at enhancing the trade-off between energy efficiency and linearity. These key technology enhancements and the evolution of MSCs within the standards will be further described in the subsequent sections.

### III. Enabling Mobile Small Cells via Network Virtualization and Softwarization

MSCs have been proposed in several studies to demonstrate the gains of solving the problem of spectrum resource allocations in the network, for instance, by offloading traffic in the congested macro-cell, and utilizing proactive caching given wireless backhaul and spectrum sharing, e.g., [6]. However, the impact of a holistic solution accommodating 4C optimization in a virtualized network has received less attention.

#### A. Virtualization Challenges

Deploying on-demand MSCs adds complexity to the virtualized network and several issues need to be addressed. Since the end-user is not receiving data from a "fixed" location, but from a "moving" transmitter, the system will need to consider spectrum, mobility and resources management for both the user and the transmitter. Network virtualization ultimately will decrease the complexity of the proposed system, enabling different RATs and MNOs to share resources, enhance resiliency, increase coverage and reduce latency. It will provide an adaptable infrastructure to the current system requirements.

In a virtualized network, physical network resources are mapped into software for enhanced resilience and lower CAPEX and OPEX. In the case of MSCs, the following 4C related problems are addressed:

- **Mobility Management (MM):** MSC deployments increase cell boundaries, potentially leading to increased HO events due to the small coverage area. Moreover, when the MSC speed is high, the HO complexity will further increase leading to higher power consumption due to excessive HO signaling. It is worth noting, that current (i.e. 3GPP LTE and NR) HO procedures are insufficient to address the new mobility challenge, providing the impetus for a complete redesign.

- **Resource Management (RM):** The RAN spectrum and infrastructure, including SCs, are envisioned to be shared between multiple 3rd party service providers, called Mobile Virtual Network Operators (MVNOs) and MRAT RANs of different ranges. Enabling this sharing requires RANs that are fully virtualized by utilizing NFV and SDN on the RAN functions and networking devices, respectively. While NFV enables RAN sharing by different tenants (MVNOs), EE Virtual Resource Management (VRM) operations of the radio and computational physical nodes require algorithms, that enable dynamic allocation and flexible management of shared resources according to Service Level Agreements (SLA) and load from each MVNO [7]. The VRM algorithms include: (i) radio resources allocations (e.g. UE-SC pairing, bandwidth allocation, UE-SC role selection and network assisted configuration, etc.); and (ii) reliable communication and computational resource allocations (e.g. NC communications, EE BBU allocations).

- **Cell offloading (CO):** Cell offloading, which in legacy 4G/low band 5G networks is based on multiple unicast sessions, is proven inefficient if co-located users request the same content. A more efficient approach would be a subgrouping system where UEs are organized in groups, and the information is sent only once to the whole group.

#### B. SECRET Approach

A virtual network usually consists of two virtual layers, called control plane and data (or user) plane. The control plane mainly handles MM, while the data plane handles communication resources. In turn, MM includes two main network procedures: HO management (both horizontal and vertical) and management of processing resources (i.e. migration of computing resources among available resource pools). In the former, an uplink reference signal (UL RS) based HO procedure proposed in [8] and depicted in Fig. 2, offers substantial benefits for MSCs. Thereby, the MSC transmits the UL RS which is simultaneously received at both serving and neighboring macro BSs. The BSs measure the UL RS and send this information to a central network controller for further measurement processing to decide which BS will serve a given MSC. Hence, measurement and further reporting transmission is not needed from the mobile device, thus reducing the HO signaling and associated power consumption.

The data plane is mainly associated with the RM, which is another core function of future CRAN. The legacy CRAN solutions, considering virtual BBU splitting, have envisioned five major logical splitting possibilities with their specific requirements (see Fig. 2). The logical sub-functions of the virtual BBUs are software-based, i.e. running in virtual machines, containers, etc. They are dynamically placed in different servers,, either in the edge data centers or embedded within the mobile users (when computational load is affordable). Such resources in the specific data centers are reserved according to the mobility characteristics of the users and the requirements for a successful and reliable communication. Communication between servers belonging to same or different data centers are empowered by network coding (NC). In particular, NC can be effectively employed as a network protocol to interconnect virtual BBUs' sub-functions, and to enhance reliability in distributed computing and storage.

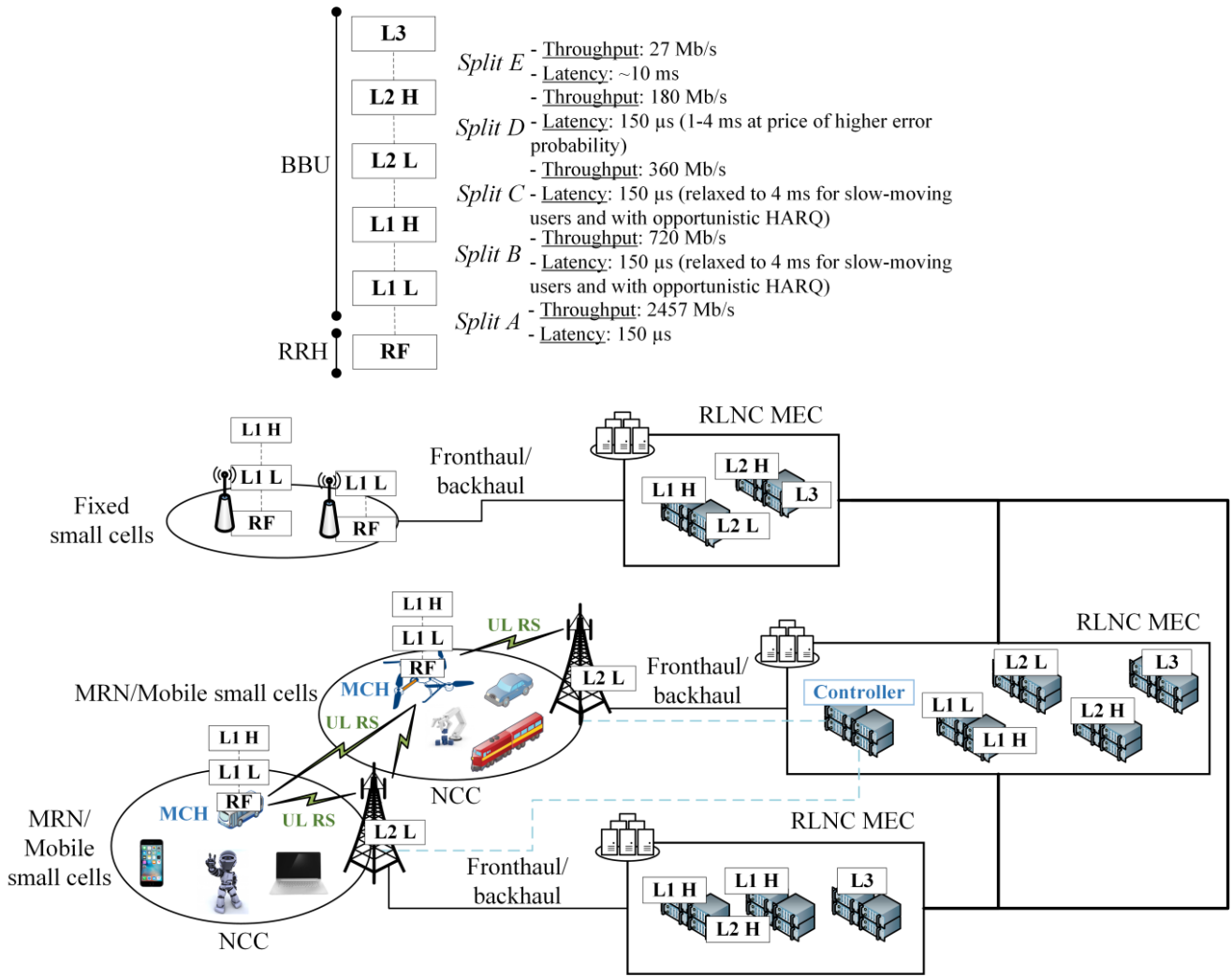

Figure 2. Representation of baseband unit logic splitting (above) and depiction of CRAN based on network-coded cooperative mobile edge computing. Small cells (fixed or mobile) communicate to edge data centers (hosting virtual baseband unit's sub-layers) via fronthaul/backhaul links, which have to satisfy the requirements highlighted above for splitting. HARQ: Hybrid Automatic Repeat Request, MRN: Mobile relay Node, RF: Radio Frequency, RLNC: Random Linear Network Coding, RRH: Remote Radio Head.

Providing a feasible deployment for NC-based subgrouping schemes requires a flexible architecture that is able to instantly deploy MSCs, where UEs with a common video request are able to collaborate locally. In this context, the SECRET communication infrastructure offers a practical cell-offloading solution that reduces the holistic energy consumption, where NCC efficiently maintains and distributes data within the MSCs by offloading cellular traffic [4]. In particular, MCHs run lower-layer virtual BBU sub-functions while upper-layer's are offloaded exploiting NCC. Such offloading can also exploit the inter-cell cooperation among different MCHs. Moreover, taking into account that short-range antennas will consume less power than macro antennas, the power consumed to transmit and receive the file is also reduced. Short range technologies (e.g., WiFi) can provide faster communication links, which will increase the overall throughput of the system. As a drawback, NCC also adds complexity and latency overhead in the network due to the coding and relaying operations. The NCC protocol comprises two different phases, the cellular phase and the cooperative phase, which can happen sequentially or in parallel. First, a cellular phase, in which the BS assigns an index to all UEs forming the cooperative cloud. It distributes the packets (previously encoded at the sender) to the mobile users via unicast TDM in a round-robin fashion. Next, a cooperative phase, in which a TDMA schedule is created and whereby each mobile user is in charge of redistributing the received packets to the remaining nodes in the cooperative cloud. Each UE will recode the received packets using the information of all packets belonging to that generation, and it will multicast the new recoded packets to the remaining cooperative nodes using short-range communication.

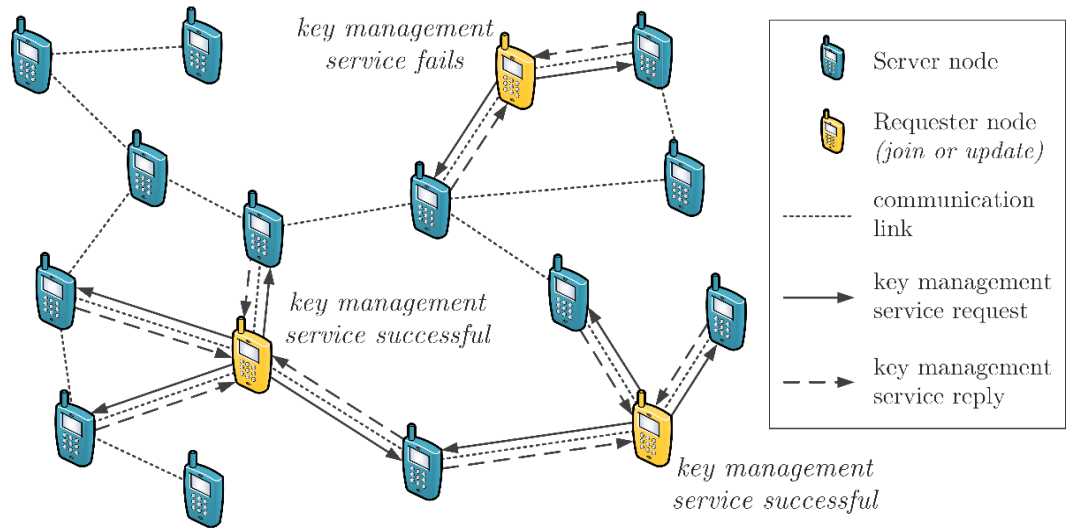

Figure 3. Illustration of a network containing 13 network nodes and 3 key management service-requesting nodes. Requester nodes require the assistance of at least (the threshold) 3 network nodes for the key management service to be successful.

IV. SECURITY PROVISIONING FOR MOBILE SMALL CELLS

In the envisioned SECRET scenarios, confidential information will be downloaded, uploaded and processed via the network of MSCs, and relayed using foreign nodes; hence, a network architecture including security is highly required.

*A. Security Challenges*

MSCs raise significant security challenges. Cryptographic security solutions are capable of solving these as long as they are supported by an efficient and robust key management (KM) scheme. A KM scheme dictates the organization of cryptographic keys to efficiently secure communication between network users. Securing MSCs implies meeting the requirements of (1) *security*, to exclude malicious users to control part of the network; (2) *connectivity*, to establish secure channels between any set of connected users; (3) *sustainability*, to provide sustained security and connectivity by means of KM; and (4) *fairness*, to evenly distribute the workload among network users and prevent selfish behavior [9].

Generally, KM schemes rely on a trustworthy and secure centralized trusted third party (TTP), but are susceptible to denial-of-service type attacks and physical compromise. Security must therefore be guaranteed using a trust-decentralizing KM scheme.

A popular solution involves certificate chains and allows any two users who wish to securely communicate but do not have any pre-existing trust relationship, to find a chain of existing trust relationships connecting each other. This solution has major security drawbacks due to the assumption that trust is both transitive and context-independent. Furthermore, users lacking a chain of trust are excluded from establishing a secure connection.

Alternative solutions involve distributing the TTP. Normally, a centralized TTP provides KM services using a master key pair. However, in a partially-distributed TTP (PD-TTP) approach, the solution relies on threshold secret sharing to split the master private key into shares that are distributed to a selected set of users, called servers. Through collaboration, they can provide the KM service. These servers are therefore burdened with the main workload. Furthermore, they may not be within transmission range of joining nodes and they may leave the network entirely over time. These issues cause a temporary or even indefinite unavailability of KM services. Finally, there is the fully-distributed TTP (FD-TTP)-based solution which is similar to the PD-TTP, except that every node receives a share of the master private key (including joining nodes). This solution is therefore capable of solving the issues present in the PD-TTP-based solution. Unfortunately, in the originally proposed solution, joining nodes are unable to verify whether the distributed KM service is trustworthy.

Therefore, a KM scheme design must satisfy all the proposed requirements and be evaluated based on its provided level of security and overheads (i.e. communication, computational and memory storage overheads).

*B. SECRET approach*

We identified that certificate chaining and PD-TTP-based solutions have inherent design flaws, whereas FD-TTP-based solutions are solvable. Therefore, our proposed DIStributed Trusted Authority-based key managemeNT (DISTANT) scheme follows the FD-TTP-based approach [10].

The DISTANT scheme adopts the self-generated-certificate public key cryptosystem for its ability to provide the highest level of trust in the FD-TTP, and the low communication overhead requirement due to non-interactive certificate updates. It consists of two phases, the network initialization phase and the operational phase.

In the network initialization phase, a TTP initiates an initial set of network users by providing them with a share of the master private key. This enables a threshold amount of them to provide KM services during network operation (cf. Fig. 3):

1. providing a requester node with its proxy key pair, enabling it to sign its self-generated certificates as if this was signed by the TTP; and

2. providing a requester node with its unique share of the master private key and join the FD-TTP group

During network operation phase, mobile nodes can join the network, self-generate certificates on-demand and exchange these self-generated certificates to establish secure communications channels.

## V. GREEN RF FOR SECRET ENABLED HANDSETS

The 5G radio transceiver front-end vision focuses on high integration level and energy efficiency in battery-powered devices. In particular, for optimizing the amount of *talk-time-per-battery-charge* in a low-voltage operation, research has focused on further increasing the average efficiency of PAs, being identified as the most power consuming RF module. The fixed supply voltage of linear amplifier classes reduces the average efficiency below 30 percent for sub-6GHz and 20 percent at mmWave frequencies, whilst the average efficiency of Green PAs is expected to improve 40 percent for acceptable 5G systems. Therefore, to meet ever increasing demands for higher data rates and wider frequency bands, novel low-cost, energy-aware and broadband handset PAs are required that mitigate thermal cooling issues and improve data throughput of MSCs.

### A. Green RF PA Challenges

The dynamic load modulation Doherty PA (DPA) is one of the dominating architectures among efficiency enhancement techniques. Due to its enhanced efficiency in deep power back-offs, the DPA has been widely used in both cellular BSs and handset devices to amplify the modulated signals with high crest factors. However, the main drawbacks of DPA implementations lie in its nonlinear distortion and intrinsic bandwidth limitation. Extensive research attempts have been conducted on improving the DPA back-off efficiency and bandwidth, covering multiband up to mmWave frequencies. Narrowband solutions attempt to improve the average efficiency without compromising its linearity, including gate bias adaptation, incorporating envelope tracking, multi-way/multi-stage and extended-resonance DPAs [11]. Although, most of these approaches yield excellent performance, harmonic tuning strategies using Class E, F, J and saturated DPAs have been identified as optimal solutions. Moreover, several efforts have been carried out on the bandwidth extension of DPAs, mostly in analog design methodologies including frequency response optimization, distributed, transformer-less, dual-input and post-matching DPAs. Nevertheless, most of the broadband DPAs have been developed for sub-6GHz low-frequency and discrete circuits and cannot be employed in IC implementation.

### B. SECRET approach

Considering the 5G user equipment, there is a need for high gain, compact size, high operation frequency and reliability. In this context, monolithic microwave integrated circuit (MMIC) technology can overcome size constraint while improving the linearity, efficiency and transducer power gain. In a fully-integrated MMIC PA all passive and active components are combined and implemented as small parts of a whole wafer to provide high device isolation and low dielectric loss. Moreover, the Gallium Arsenide (GaAs) is a preferred material at mmWave frequencies mainly due to its high electron mobility that implies a greater electron velocity for a given electric field. In fact, GaAs FET (Field Effect Transistor) is a mature technology with a high cost advantage that has led to widespread adoption in the cellular communication market. Table I provides the performance comparison of recently reported handset DPAs based on the semiconductor technologies [12]. Compared to state-of-the-art and published experimental results, the GaAs pHEMT PAs deliver excellent performances in terms of output power and back-off efficiency. In SECRET, we adopted the 0.25-μm InGaAs/GaAs (E-mode) pHEMT technology that is ideally suited for 5G handset PAs with positive supply voltage to provide an ultra-compact MMIC DPA. The proposed symmetrical design technique is based on the wideband Class-

TABLE I. PERFORMANCE COMPARISON OF HANDSET DPAS

| Technology[a] | Freq. (GHz) | Pout (dBm) | Gain (dB) | PAE BO(%) | Class | Area $mm^2$ |
|---|---|---|---|---|---|---|
| 28-nm CMOS | 32 | 19.8 | 22 | 12.8 | AB/C | 1.79 |
| 0.13-μm SiGe | 28 | 16.3 | 18.2 | 13.9 | B/C | 1.76 |
| 180-nm SiGe BiCMOS | 24-30 | 28 | 20 | 20 | AB/AB | 4.19 |
| 2-μm InGaP/GaAsHBT | 2.5-2.7 | 24.6 | 19 | 25 | AB/C | 1.44 |
| D-mode GaAs | 23-25 | 30 | 12.5 | 20 | B/C | 4.29 |
| E-mode 0.15-μm GaAs | 28 | 26 | 12 | 29 | AB/B | 2.85 |
| E-mode 0.15-μm GaAs | 25-29 | 28.2 | 15 | 27 | AB/C | 4.93 |
| E-mode 0.25-μm GaAs | 9-10.8 | 29.2 | 12.2 | 50 | J/J | 3.15 |

[a.] A description and references to the listed techniques can be found in [12].

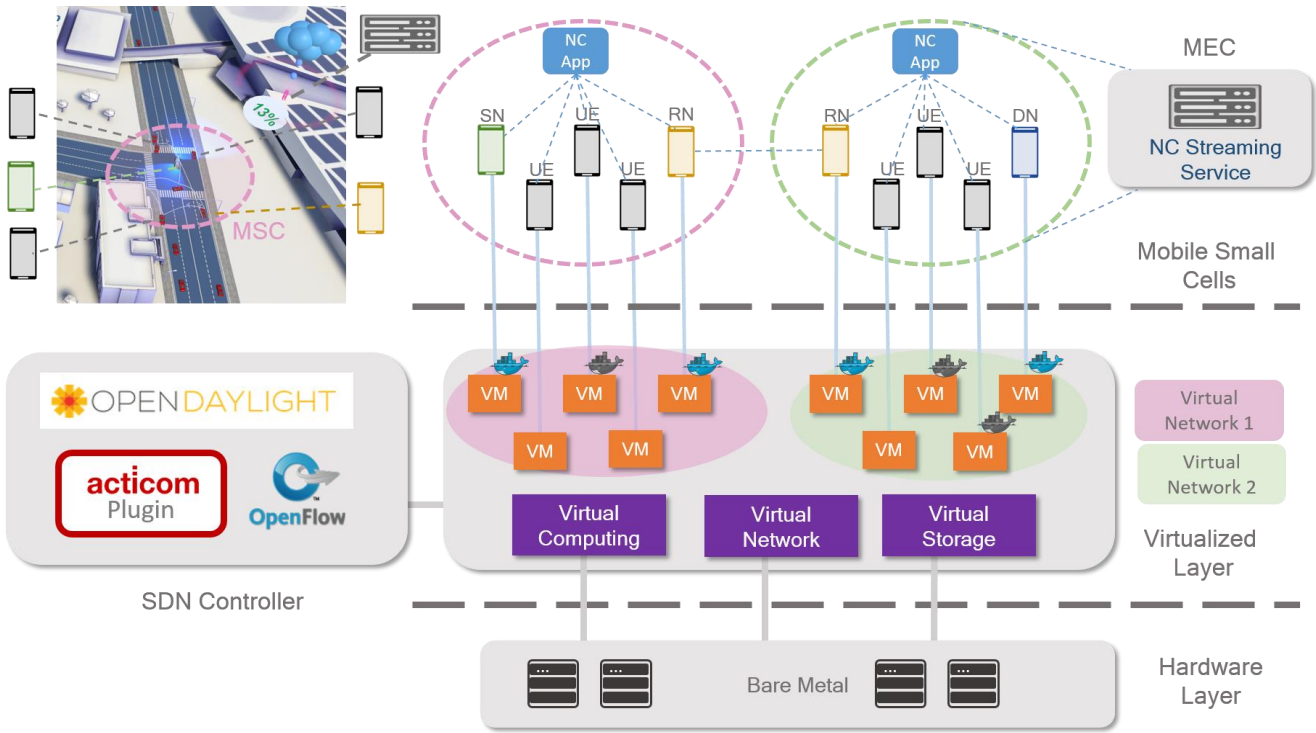

Figure 4. SECRET testbed architecture.

J mode of amplification. The harmonic generating Class-J PA exploits the capacitive second harmonic load components to absorb the parasitics that can increase the fundamental voltage and maximize the efficiency. Besides, the employed broadband post-matching network reduces the impedance transformation ratio of the conventional DPA and restores the proper load modulation.

VI. DEMONSTRATING MOBILE SMALL CELLS - SECRET TESTBED

The SECRET approach has been validated by means of a testbed that is illustrated by Fig. 4. The testbed has two well-differentiated layers, namely a hardware layer and a virtualized layer. The testbed uses SDN in the virtualized layer with two different observable planes, namely, the data plane and the control plane.

The hardware layer consists of an *HP Prodesk 600 G1* that emulates different cellular eNBs, and three *Fujistu MPC3s* that create virtualized MSC (vMSC) and the UEs inside them. They are all connected to the Internet via a *TP-Link C9* router.

The control plane of the virtualized layer is programmable and managed by an SDN controller. The key components of the testbed are the Openstack cloud platform to manage network functions, and OpenDayLight SDN controller to manage the network flows. In the setup, the orchestrator utilizes OpenStack APIs to start virtualized MSC functions and instructs the SDN controller to manage the flows. The virtual machines (VM) are spawned on a compute node that acts as UEs (virtual objects). Kernel-based Virtual Machine (KVM), which is a full virtualization solution for Linux on x86 hardware, is used as a hypervisor on the compute node to host virtual machines. In the testbed, we consider a MSC that is deployed as an overlay network. MEC server is assumed to be deployed at the eNB and the NC streaming service runs in the MEC. The proposed approach is based on an overlay network that logically interconnects all the participating UEs in the MSC to the physical network. Each virtual UE is identified by a MAC address to enable the communication among the virtual UEs in the overlay network. The SDN-based network allows isolation of the overlay network through OpenFlow rules.

The data plane of the virtualized network consists of a NCC streaming service, which is placed in the MEC, multiple MSCs, whose controller is placed at the MCH, and NCC apps, which are placed in the UEs. A video is streamed from the server to the UEs to showcase the traffic offload and the increased throughput.

In the demonstrator, a MSC is placed inside an ambulance, which works as the MCH. A docker container runs in the nodes to show whether the node is active, inactive, or in stand-by. The MSC in the ambulance offloads the cellular traffic achieving a 13 percent cellular utilization (cf., Fig. 4) and acts as an intermediate node between surrounding cars and the eNB. Running video streaming could demonstrate hours of reliable communication with a decoding ratio over 99.5 percent, which can increase further by increasing the number of coded transmissions in the MSC. NCC-assisted video can be beneficially used to provide services with demands on throughput, latency and reliability, e.g., tele-surgery monitoring.

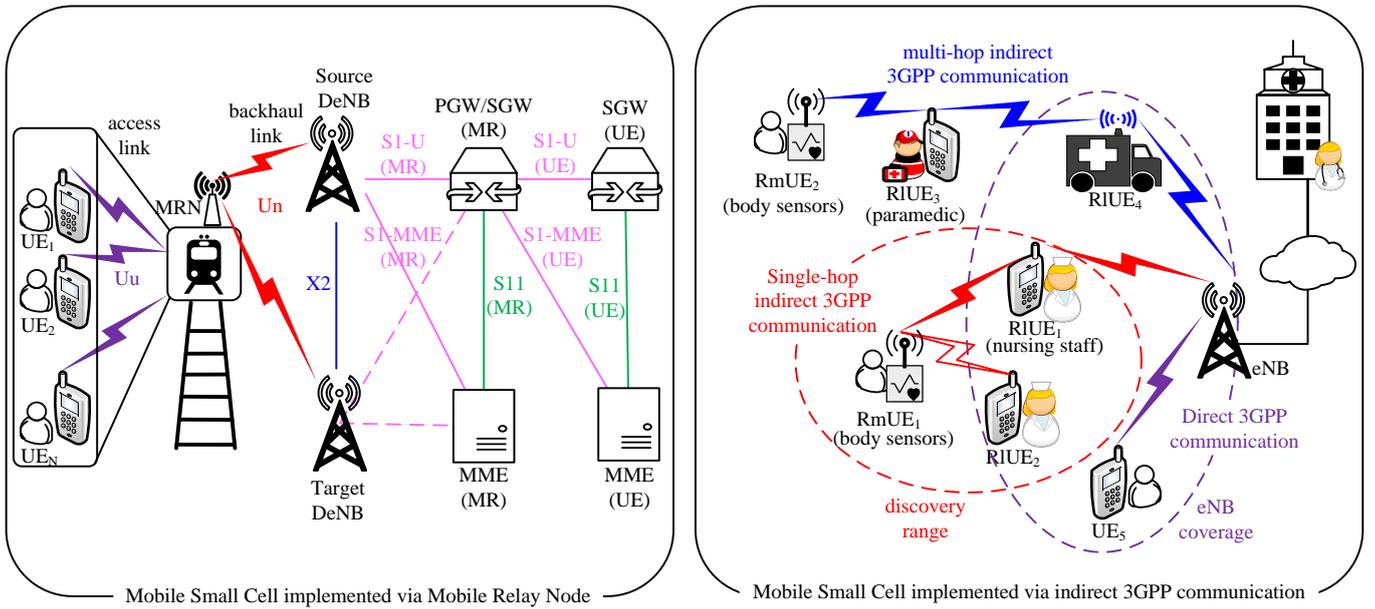

Figure 5. Mobile relay architecture (see Alt. 1 [15]), supporting inter-DeNB handovers (HOs) (left) and the realization of small cells via indirect 3GPP communication (right).

VII. STANDARDIZATION TOWARDS MOBILE SMALL CELLS

The realization of the MSC paradigm, has lately received attention by 3GPP, where different technologies have been proposed for potential inclusion in the standards. Notably, the study of Mobile Relay (MR) technologies was undertaken in Release 11 (R11) and R12. Moreover, the 3GPP standard also includes D2D proximity services (ProSe), enabling so-called 3GPP Indirect Communication. The use of these technologies towards the implementation of MSCs will be addressed hereafter along with standard evolution.

*A. Mobile Relays*

Originally targeting high-speed trains, MRs aim at improving the cell coverage of on-board UEs by deploying access devices, namely Mobile Relay Nodes (MRN), providing both a wireless backhaul connection via the eNBs alongside the vehicle's trajectory, as well as wireless connectivity to UEs inside the vehicle. The concept may be easily extended to include more general types of transportation, sharing some similar scenario characteristics, namely: moderate to high speeds; known (or predictable) trajectory in most cases; high penetration loss through the vehicle cabin; and UEs being stationary or semi-stationary w.r.t. the vehicle's movement.

The MRN is wirelessly connected to a Donor eNB (DeNB) via the Un radio interface, while on-board UEs are connected to the MRN via the Uu interface. A MRN supports a subset of the UE functionalities and the eNB functionalities in order to connect to the DeNB. In alignment with SECRET's MSC, MRNs also support multi-RAT functionalities by utilizing LTE Un as a backhaul link and different air interface technologies (i.e LTE/3G/2G/WiFi) on the access link. When the Un interface connection is changed to a different DeNB, via an inter-DeNB HO, MRNs should still provide uninterrupted connectivity to the on-board UEs towards the CN.

Overall, MRN technology is a suitable candidate for accommodating the functionality of MSCs in 3GPP B5G networks. Noteworthy, in [13] several MR architectures were proposed, Alt.1 to Alt. 4, of which, based on a comparative study therein, Alt.1 and Alt.2 were selected for further work on MRs. Figure 5 (left) illustrates the Alt. 1 architecture, where the MR HO reuses existing UE HO procedures with some modifications.

*B. Indirect 3GPP Communication*

From R13 onwards, the 3GPP standard may support the realization of the MSC concept using D2D proximity services (ProSe), allowing a two-hop UE-to-Network relay option, coined as Indirect 3GPP Communication (i3Com). Although initially intended to support emergency responders during public safety and security situations, further scenarios have been identified in R16 and beyond, e.g., inHome, SmartFactories, Logistics, etc.

I3Com shall be supported between a Remote UE (RmUE) and the network via a Relay UE (RlUE), where the connection between the UEs shall be able to use E-UTRA or WLAN [14]. The maximum number of RlUEs between an RmUE and the network is one (single-hop) in R16, but may be relaxed in R17 allowing multi-hop [15]. The network will be responsible for authorizing, enabling and disabling a UE to act as RlUE. The control of the RmUE by the network will also be done via i3Com using either E-UTRA or WLAN. Several RlUEs may be available in the proximity of the RmUE, and thus i3Com shall support discovery, selection and reselection of an RlUE based on a combination of different criteria [15]. Reselection may be triggered by any dynamic change in the selection criteria, e.g. by the battery of a RlUE getting depleted, a new relay-capable UE getting in range, a RmUE requesting higher QoS, etc.

Generally, the use of i3Com should typically not lead to an increase in power consumption at the RmUE when compared to direct 3GPP communication for the same traffic. Figure 5 (right) illustrates the concept of i3Com.

*C. Future standardization efforts enabling small cells*

Lately, the resurgence of relay technology seems to have picked momentum in 3GPP with the introduction of Integrated Access and Backhaul (IAB) technology. The deployment of IAB nodes enable spectrum pooling between the access link and the backhaul link on a per-demand basis. It is expected that the current stationary definition of IAB nodes will evolve into mobile IAB concepts in future releases, thus enabling the MCS concept.

Another anticipated direction for future 3GPP standards is the promotion of a user-centric cell-free connection approach that will enable power-efficient mobility management. An uplink sounding reference signal (SRS) based HO procedure proposed in SECRET [8] reveals improved performance in terms of HO signaling overhead in SC deployments. Extending this concept to MSCs or the MRNs is expected to bring similar gains.

## VIII. CONCLUSIONS

This paper presents the SECRET approach towards B5G/6G networking based on virtual mobile small cells. These are perceived by the network as a pool of new networking resources inheriting 4C capability, which can be set-up on-demand enabling new 6G compliant services, such as URLLC, as well as significantly reducing the cost of infrastructure-based deployments.

Towards this end, challenges, enabling technologies and solutions have been discussed together with a testbed demonstrating the SECRET proof-of-concept. Controlled by a distributed SDN architecture, it has been shown that the SECRET architecture enables reliable and efficient connectivity, with inherent capability to support low latency and cloud computing 6G services going beyond the edge node. Finally, the evolution of such an approach in future standards is foreseen.

REFERENCES

[1] M. Latva-aho, and K. Leppänen, "Key drivers and research challenges for 6G ubiquitous wireless intelligence," 6G Research Visions 1, Sept. 2019 (white paper); ISBN 978-952-62-2354-4 (online).

[2] SECRET project (H2020 MCSA-ETN), project no: 722424; http://h2020-secret.eu/.

[3] F. Rebecchi, *et al*., "Data Offloading Techniques in Cellular Networks: A Survey," in *IEEE Communications Surveys & Tutorials*, vol. 17, no. 2, pp. 580-603, Secondquarter 2015.

[4] S. Irum, *et al*. "Network-coded Cooperative Communication in Virtualized Mobile Small Cells," *2019 IEEE 2nd 5GWF*, Dresden, Germany, 2019, pp. 264-268

[5] H. Ning, *et al*., "A Survey and Tutorial on "Connection Exploding Meets Efficient Communication" in the Internet of Things," in *IEEE Internet of Tings Journal*, vol. 7, no. 11, pp. 10733-10744, Nov. 2020.

[6] Y. M. Kwon *et al*., "Performance Evaluation of Moving Small-Cell Network with Proactive Cache," *Mobile Information Systems*, vol. 2016, 2016.

[7] F. Marzouk, J. P. Barraca and A. Radwan, "On Energy Efficient Resource Allocation in Shared RANs: Survey and Qualitative Analysis," in *IEEE Communications Surveys & Tutorials*, vol. 22, no. 3, pp. 1515-1538, thirdquarter 2020.

[8] M. Tayyab, *et al*., "Uplink Reference Signals for Energy-Efficient Handover," in *IEEE Access*, vol. 8, pp. 163060-163076, 2020.

[9] M. De Ree, *et al*., "Key Management for Beyond 5G Mobile Small Cells: A Survey," in *IEEE Access*, vol. 7, pp. 59200-59236.

[10] M. de Ree, *et al*., "Distributed Trusted Authority-based Key Management for Beyond 5G Network Coding-enabled Mobile Small Cells," *2019 IEEE 2nd 5GWF*, Dresden, Germany, pp. 80-85, 2019.

[11] M. Sajedin, *et al*., "A Survey on RF and Microwave Doherty Power Amplifier for Mobile Handset Applications," *Electronics*, 2019, vol. 8, no. 6, pp.1-31.

[12] W. Chen, *et al*., "Doherty PA for massive MIMO," *IEEE Microwave magazine*. 2020, 1527-3342/20:78-93.

[13] 3GPP TR 36.836 V12.0.0. "Study on mobile relay (Release 12)," June 2014.

[14] 3GPP TS 22.278 V17.1.0. "Service requirements for the Evolved Packet System (EPS) (Release 17)," Dec. 2019.

[15] 3GPP TS 38.300 V15.0.0. "NR; NR and NG-RAN Overall Description; Stage 2 (Release 15), " Dec. 2017.